\documentclass[conference]{IEEEtran}
\usepackage{graphicx}
\usepackage{pgf}
\usepackage{tikz}
\usetikzlibrary{arrows,automata}
\usepackage[latin1]{inputenc}
\usepackage{subcaption}
\usepackage{amssymb}
\usepackage{tabularx,booktabs}
\usepackage{pgfplots}
\usepackage{lipsum}
\usepackage{subcaption}
\pgfplotsset{compat=1.6}
\usepackage{amsmath}
\usepackage{mathtools}
\usepackage{cleveref}

\DeclarePairedDelimiter\floor{\lfloor}{\rfloor}
\ifCLASSINFOpdf
\else
\fi
\makeatother
\makeatletter
\newcommand{\linebreakand}{%
\end{@IEEEauthorhalign}
\hfill\mbox{}\par
\mbox{}\hfill\begin{@IEEEauthorhalign}
}
\makeatother

\begin{document}
	\title{Translating Cyber-Physical Control Application Requirements to Network level Parameters}
	\thanks{This work was supported by Swedish Foundation for Strategic Research (SSF) under Grant iPhD:ID17-0079.}
	\author{\IEEEauthorblockN{Milad Ganjalizadeh\IEEEauthorrefmark{1}\IEEEauthorrefmark{2},
			Abdulrahman Alabbasi\IEEEauthorrefmark{1}, Joachim Sachs\IEEEauthorrefmark{1}, and
			Marina Petrova\IEEEauthorrefmark{2}}
		\IEEEauthorblockA{\IEEEauthorrefmark{1}Ericsson Research, Ericsson AB, Stockholm, Sweden\\
			\IEEEauthorrefmark{2}School of Electrical Engineering and Computer Science, KTH Royal Institute of Technology, Stockholm, Sweden \\
			Email: \IEEEauthorrefmark{1}\{milad.ganjalizadeh, abdulrahman.alabbasi, joachim.sachs\}@ericsson.com, 
			\IEEEauthorrefmark{2}petrovam@kth.se}
	}
	
	\maketitle
	
	\begin{abstract}
		Cyber-physical control applications impose strict requirements on the reliability and latency of the underlying communication system. Hence, they have been mostly implemented using wired channels where the communication service is highly predictable. Nevertheless, fulfilling such stringent demands is envisioned with the fifth generation of mobile networks (5G). The requirements of such applications are often defined on the application layer. However, cyber-physical control applications can usually tolerate sparse packet loss, and therefore it is not at all obvious what configurations and settings these application level requirements impose on the underlying wireless network. In this paper, we apply the fundamental metrics from reliability literature to wireless communications and derive a mapping function between application level requirements and network level parameters for those metrics under deterministic arrivals. Our mapping function enables network designers to realize the end-to-end performance (as the target application observes it). It provides insights to the network controller to either enable more reliability enhancement features (e.g., repetition), if the metrics are below requirements, or to enable features increasing network utilization, otherwise. We evaluate our theoretical results by realistic and detailed simulations of a factory automation scenario. Our simulation results confirm the viability of the theoretical framework under various burst error tolerance and load conditions.
	\end{abstract}
	
	\IEEEpeerreviewmaketitle
	
	\section{Introduction}
	\IEEEPARstart{C}{yber}-Physical Systems (CPS) are engineered systems with integrated control, networking, and computing infrastructures that can interact with the physical world. This joint emphasis on the physical process, communication, and computation in CPS is a key enabler for the development of future technologies \cite{Baheti2011}. From the communication perspective, the tight interaction of the CPS with the physical environment put stringent requirements on the observed quality of service (QoS). The next generation wireless network, 5G, is being designed to fulfill strict latency and reliability requirements, and support ultra-reliable and low-latency communications (URLLC). Hence, 5G has the potential to assist cyber-physical applications \cite{xu2018}.
	
	CPS applications will revolutionize our society in many ways, such as the industry and healthcare sectors. These include process and factory automation, remote surgery, advanced automotive systems, and smart electric power distribution. Such applications are mostly designed to control the physical process and hence, are referred to as cyber-physical control applications. Wireless communication services supporting the aforementioned applications are required to be ultra-reliable (e.g., mean time to failure of more than one year) with high levels of service availability (e.g., $99.9999\%$ successful service delivery) \cite{3GPP22104}.
	
	Within the context of 5G, the terms "reliability" and "availability" are often used interchangeably, whilst the definition varies \cite{HoblerCommL2018}. Reliability metrics are traditionally defined on the physical (PHY) layer, and indicators, such as outage probability, bit error ratio (BER), or block error ratio (BLER), are used to identify the QoS of the channel (e.g., \cite{Zhang1999}). In general, the percentage of successful transmissions seems to be used as an indicator for the observed reliability on both the PHY and link layer. This implies that, with such an indicator, link layer reliability enhancement techniques (e.g., medium access control (MAC) layer retransmissions) are as well taken into consideration (e.g., \cite{Anand2018}). Nevertheless, in wireless communications, errors often occur in bursts, and such metrics cannot capture the impact of the correlated failures on the performance. For instance, the analysis in \cite{Ganjalizadeh2019} and \cite{nielsen2019reliability} show that the impact of space-time correlation on the end-to-end performance is considerable.
	
	In reliability theory \cite{Rausand2004}, there is an indicator that keeps track of the duration of consecutive failures, which has been widely used in wireless communications to model the correlation among failures. In the 60s, Gilbert, in \cite{Gilbert1960}, and Elliot, in \cite{Elliott1963}, proposed a two-state Markov model to analyze bursty errors in wireless channels. One of the recent applications of their model is to derive reliability metrics for multi-connectivity scenarios using the average fade duration in Rayleigh fading channels (\cite{HoblerCommL2018} and \cite{HoBlerGlobecom2019}). In \cite{Ohmann2015}, a threshold on fade duration is introduced to model reliability enhancement techniques such as coding and retransmissions. The assumption is that the outages shorter than this threshold can be tolerated due to PHY or link layer reliability enhancement techniques. Nevertheless, according to \cite{3GPP22104}, the requirements for cyber-physical control applications are defined on the application level. Thus, there is a need for a shift in the focus of research on wireless communication systems that take into account application level performance. As an example, consider the case that new positions are sent periodically to mobile robots. In this case, a limited number of consecutive packet loss will not impact the overall performance of the robot. In other words, when a new packet is successfully received, the application layer can use methods, such as interpolation, to estimate the lost positions. This time period for which the application layer can tolerate failures in a wireless network is called survival time. Therefore, application level requirements cannot be directly used as the expected network performance. To the best of our knowledge, this is the first article that relates the application level performance of CPS to network level parameters.
	
	Motivated by the above, and focusing on time domain aspects of wireless communication systems, we make an effort to bridge the gap between traditional network level performance metrics (e.g., packet error ratio) and application level performance metrics (e.g., availability). Therefore, we propose a finite-state Markov chain (FSMC) to keep track of burst errors and get an estimate of the error length distributions. The proposed framework aims to provide a mapping function that takes the application's survival time into account and maps the requirements on the application layer to the expected network performance, or vice versa. Moreover, we perform simulations to verify and evaluate our mapping function in a 3GPP inspired factory automation use case.
	
	The remainder of this paper is organized as follows. We describe our system model in Section II, and define network level and application level reliability quantities in Section III. The mapping function connecting the reliability quantities between two layers is derived in Section IV. Section V evaluates the precision of our mapping function on a simulation performed for a factory automation scenario, and Section VI summarizes concluding remarks.
	\vspace{-0.1em}
	\section{System Model}
\begin{figure}
		\centering
		\includegraphics[width=0.9\columnwidth]{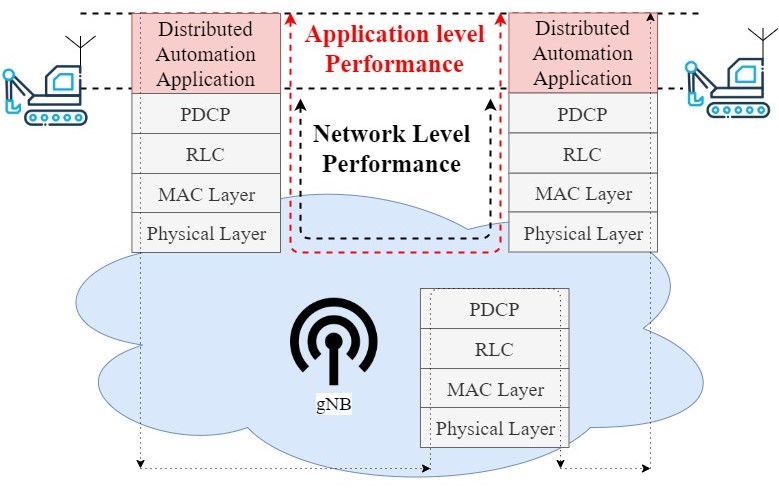}
		\vspace{-.5em}
		\caption{System model}
		\label{fig:systemModel}
		\vspace{-1.5em}
	\end{figure}

	Cyber-physical control applications consist of end nodes performing a variety of functions, all of which contribute towards controlling physical objects. Examples of such end nodes are sensors, actuators, and switches. In such a system, wireless communication is responsible for end-to-end message delivery between these nodes (i.e., from the application layer of the source node to the application layer of the destination node). Once the messages are delivered, the application layer performs the requested function. However, the operations within a wireless network can be interrupted due to packet loss or delay for reasons such as interference, fading, congestion, insufficient signal power, or hardware and software bugs. Recall that a single packet loss on wireless networks will not necessarily affect the performance observed by the application. Hence, we discuss the performance of CPS on two levels, as observed by: (i) the application, and (ii) the network. The former can be considered as the end-to-end performance, while the latter is the performance that network can provide up to the packet data convergence protocol (PDCP) layer. \figurename\,\ref{fig:systemModel} illustrates different performance levels with the complete protocol stack. This figure does not take fixed network parts, such as core network, into consideration. In this paper, we focus on single-user performance, where we address the mapping between network level and application level performance.
	
	Consider a wireless communication system where packets are being transmitted and expected to be received by an end node, illustrated in \figurename\,\ref{fig:systemModel}. The network is considered to be in up state, \(\mathcal{U}_N\), if packets are correctly received within the target delay bound. When the first packet is lost or received after its latency deadline, the network transits to the down state,  \(\mathcal{D}_N\). The network transits back to \(\mathcal{U}_N\) with the first successfully received packet. From the application perspective, there exists a time period that consecutive failures in the communication system can be tolerated. This time period is called survival time, \( T_{sv}\), and application layer stays in up state, \(\mathcal{U}\), during this period while the network is down. If failures in communication persist for a period longer than \(T_{sv}\), the application also transits to down state (\(\mathcal{D}\)). While both network and application are in down state, successful reception of a packet leads to transition of both network and application layer to up state, \(\mathcal{U}_N\) and \(\mathcal{U}\), respectively.
	
	According to \cite{3GPP22104}, most of the traffic for cyber-physical control use cases is cyclic. In cyclic traffic, a certain amount of application data is transmitted periodically, i.e., the time difference between two consecutive transmissions is deterministic. In this case, for cyclic traffic with period \(T_c\), the survival time, \(T_{sv}\), can be derived as the number of consecutive packet failures that can be tolerated by the application as
	\begin{eqnarray}
	N_{sv}=\floor*{\frac{T_{sv}}{T_c}}.
	\end{eqnarray}
	
	\section{Reliability Quantities} \label{Reliability_Quantities}
	In literature, the reliability and resilience of a repairable system are characterized by different temporal performance metrics, where a system can consist of a single or large number of components \cite{Rausand2004}. We consider the end nodes as repairable items and apply the reliability theory definitions to wireless  systems to address the reliability and resilience of cyber-physical control applications. Note that all of the definitions below comply with the 3GPP standard defined in \cite{3GPP22104}.
	\vspace{-0.4em}
	\subsection{Reliability}
	\vspace{-0.4em}
	In reliability literature, the term reliability is defined as the mean duration of time that the system is operational, under stated requirements. Therefore, application level reliability, \(R\), and network level reliability, \(\bar{\tau}_{\mathcal{U}_N}\), can be defined as the mean time for which the communication system is in state \(\mathcal{U}\) and \(\mathcal{U}_N\), respectively. Sometimes, in \cite{Kuo2003} for example, the reliability is described as the mean time between failures (MTBF) with conceptually the same meaning as our definition.
	\vspace{-0.4em}
	\subsection{Mean Down Time}
	\vspace{-0.4em}
	We define the mean down time as the mean duration of time that the system is not performing its required functions under stated conditions. On application level, it is the mean time that the communication system is in state \(\mathcal{D}\) and is denoted as \(\bar{\tau}_{\mathcal{D}}\). On network level, it is the mean time that communication system is in state \(\mathcal{D}_N\) and is denoted as \(\bar{\tau}_{\mathcal{D}_N}\). In some literature (e.g., \cite{Kuo2003}, \cite{Ohmann2015}), the mean down time is addressed as the mean time to repair (MTTR).
	\vspace{-0.4em}
	\subsection{Availability}
	\vspace{-0.4em}
	Availability is the ability of an item to perform its functions as expected at an arbitrary instant of time \cite{Rausand2004}. Intuitively, a communication system availability (or average availability) can be defined as the mean proportion of time for which the system is in up state, i.e., communication service is delivered according to the agreed QoS \footnote{It is worth noting that some of the work addressing lower layer performance of wireless communication systems, use the term reliability (instead of availability) for the same definition as above (e.g., \cite{Nielsen2018}).}. The network level availability is denoted as \( A_{N}\) and can be calculated as
	\begin{eqnarray}\label{A_N_def}
	A_N=\frac{\bar{\tau}_{\mathcal{U}_N}}{\bar{\tau}_{\mathcal{U}_N}+\bar{\tau}_{\mathcal{D}_N}}.
	\end{eqnarray}
	Assuming cyclic traffic with period, \(T_c\), network level availability can be derived based on the expected packet error ratio, \(p\), as
	\begin{equation}
	\label{A_N}
	\begin{split}
	A_N = 1-\frac{\bar{\tau}_{\mathcal{D}_N}}{\bar{\tau}_{\mathcal{U}_N}+\bar{\tau}_{\mathcal{D}_N}}&=\lim\limits_{T \to \infty}\frac{\sum_{i\in T}{\tau_{\mathcal{D}_{N,i}}}}{\sum_{i\in T}{\tau_{\mathcal{U}_{N,i}}+\tau_{\mathcal{D}_{N,i}}}} \\ 
	&=  1-\frac{N_f T_c}{N T_c}=1-p,
	\end{split}
	\end{equation}
	where \({\tau_{\mathcal{U}_{N,i}}}\) and \({\tau_{\mathcal{D}_{N,i}}}\) represent the \(i\)th observed time period for which there are consecutive successful receptions and consecutive failures, respectively. Moreover, \(N_f\) and \(N\) are the number of packet failure and total number of packet transmissions, respectively, over the period of \(T\). 
	
	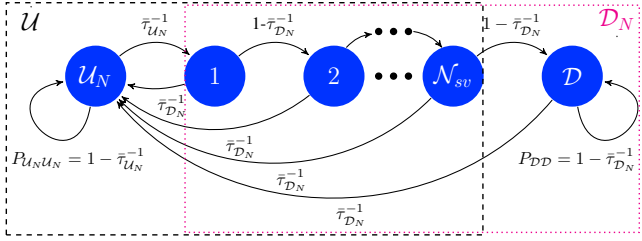
\begin{figure}
		\centering
		\scalebox{0.66}{\definecolor{aqua}{rgb}{0.0, 1.0, 1.0}
\begin{tikzpicture}[->,>=stealth',shorten >=1pt,auto,node distance=2.4cm,
semithick]
	\tikzstyle{every state}=[fill=blue!80!cyan,draw=none,text=white,minimum size=1.23cm]

		\node[state] (A)              {\Large$\mathcal{U}_N$};
		\node[state]         (B) [right of=A] {\Large $1$};
		\node[state]         (C) [right of=B] {\Large$2$};
		\node[state]         (D) [right of=C] {\Large$\mathcal{N}_{sv}$};
		\node[state]         (E) [right of=D] {\Large$\mathcal{D}$};

   \path (A) edge [out=260,in=190, looseness=6] node {} (A)
			  edge [bend left=40]  node {$\bar{\tau}^{-1}_{\mathcal{U}_N}$} (B)
	     (B) edge [bend left=40]  node {1-$\bar{\tau}^{-1}_{\mathcal{D}_N}$} (C)
		     edge [bend left=15]  node {\quad\quad$\bar{\tau}^{-1}_{\mathcal{D}_N}$} (A)
	     (C) edge [bend left=20]  (5.6,0.85)
			 edge [bend left=38]  node {\quad\quad\quad$\bar{\tau}^{-1}_{\mathcal{D}_N}$} (A)
	     (D) edge [bend left=40]  node {$1-\bar{\tau}^{-1}_{\mathcal{D}_N}$} (E)
		     edge [bend left=45]  node {\quad\quad$\bar{\tau}^{-1}_{\mathcal{D}_N}$} (A)
		 (E) edge [bend left=50]  node {\quad\quad$\bar{\tau}^{-1}_{\mathcal{D}_N}$} (A)
		     edge [out=280,in=-10, looseness=6] node {} (E);
\draw
(6.4,0.9) to [out=-10,in=130](7,0.55);
\node at (-.35,-1.68) {$P_{\mathcal{U}_N\mathcal{U}_N}=1-\bar{\tau}^{-1}_{\mathcal{U}_N}$};
\node at (9.7,-1.68) {$P_{\mathcal{D}\mathcal{D}}=1-\bar{\tau}^{-1}_{\mathcal{D}_N}$};
\foreach \i in {1,...,3}
{
	\filldraw[color=black, fill=black,  thick](6.6-0.3*\i,0) circle (0.06);
	\filldraw[color=black, fill=black,  thick](6.6-0.3*\i,0.9) circle (0.06); 
}
\draw[black, thick, dashed] ((-1.8cm,-3.2cm) rectangle (7.8cm,1.48cm);
\filldraw 
(-1.3cm,1.4cm) circle (0.1pt) node[align=right,   below] {\color{black}\Large$\mathcal{U}$};

\draw[magenta, thick, dotted] ((1.8cm,-3.15cm) rectangle (10.95cm,1.43cm);
\filldraw 
(10.5cm,0.8cm) circle (0.1pt) node[align=right,  above] {\color{magenta} \Large$\mathcal{D}_N$};
\end{tikzpicture}}
		\caption{Proposed Markov chain to keep track of burst length}
		\label{fig:markovChain}
		\vspace{-1.5em}
	\end{figure}

	\section{Mapping of reliability quantities between different layers}
	
	The 3GPP technical specification in \cite{3GPP22104} specifies not only the required application level reliability and availability but also typical survival time for different use cases within cyber-physical control applications. Our mapping function allows operators to derive the exact requirements on PDCP layer given their end-to-end requirements or, alternatively, obtain the performance as application observes it.
	
	To derive the mapping between different layers, we propose an FSMC to keep track of burst error length on the communication system. The state space is partitioned as 
	\begin{eqnarray}
	\mathcal{S}=\{\mathcal{U}_N, 1, 2, ..., \mathcal{N}_{sv}, \mathcal{D}\},
	\end{eqnarray}
	where states \{\(1, 2, \cdots, \mathcal{N}_{sv}\)\} represent the consecutive packet failures with length of \(1, 2, \cdots, N_{sv}\), respectively. Therefore, the subset of state space \{\(1, 2, \cdots, \mathcal{N}_{sv}, \mathcal{D}\)\} and \{\(\mathcal{U}_N, 1, 2, \cdots, \mathcal{N}_{sv}\)\} are representing network level \(\mathcal{D}_N\) and application level \(\mathcal{U}\), respectively. However, the states are broken down to capture burst length up to the size of survival time, and consequently to have a proper representation of different layers. In other words, the network is available if the communication system is in state \(\mathcal{U}_N\), whilst the application is available if the communication system is in the subspace \{\(\mathcal{U}_N, 1, 2, ..., \mathcal{N}_{sv}\)\}.
	
	In this model, the steady-state probability of each state represents the long-term time proportion that system spends in that specific state. Therefore, using \eqref{A_N}, the steady-state probability of \(\mathcal{U}_N\) is
	\begin{eqnarray}
	\label{pi_Un}
	\pi_{\mathcal{U}_N}=A_N=1-p .
	\end{eqnarray}
	On the other hand, the steady-state probability of \(\mathcal{D}\), $\pi_\mathcal{D}$, as the application level unavailability (i.e., \(1-A\)) of the receiving node, is usually given as a requirement. \figurename\,\ref{fig:markovChain} shows the proposed FSMC and the transition probabilities for different states. Let us denote the transition probability from state \(i\) to state \(j\) by \(P_{ij}\) where \(i,j\in\mathcal{S} \). Since the mean time that the system stays in state \(\mathcal{U}_N\) is \(\bar{\tau}_{\mathcal{U}_N}\), then
	\begin{eqnarray}
	P_{\mathcal{U}_N\mathcal{U}_N}=1-\bar{\tau}^{-1}_{\mathcal{U}_N}.
	\end{eqnarray}
	On the other hand, all the \(P_{ij}\)  where \(j=i+1\) and \(i\in\{1, 2, ..., \mathcal{N}_{sv}\}\) are the same and equal to the probability that communication system stays in \(\mathcal{D}_N\). 
	
	A Markov chain is irreducible if all of its states communicate with each other. The period of a state is the greatest common divisor of all integers $m$, such that the probability of returning to that same state in $m$ transitions is greater than 0. Consequently, a state with period of $1$ is called aperiodic. Since the proposed FSMC is irreducible and all of its states are aperiodic, it has a unique stationary distribution (proof in \cite{papoulis2002}). Thus, the transition matrix \(\mathbf{M}\) can be derived as
	
	\[\mathbf{M}=
	\begin{bmatrix}
	1-\bar{\tau}^{-1}_{\mathcal{U}_N} & \bar{\tau}^{-1}_{\mathcal{U}_N} & 0 & \dots  & 0 \\
	\bar{\tau}^{-1}_{\mathcal{D}_N} & 0 & 1-\bar{\tau}^{-1}_{\mathcal{D}_N} & \dots  & 0 \\
	\vdots & \vdots & \vdots & \ddots & \vdots \\
	\bar{\tau}^{-1}_{\mathcal{D}_N} & 0 & 0 & \dots  & 1-\bar{\tau}^{-1}_{\mathcal{D}_N}
	\end{bmatrix},
	\]
	where \(\mathbf{M}\) is an square matrix with \(N_{sv}+2\) rows and columns. Let us denote the array of steady-state probabilities by \(\overrightarrow{\pi}\). Using Markov chain properties we have
	\begin{eqnarray}\label{markov_prop1}
	\overrightarrow{\pi}\times\mathbf{M}=\overrightarrow{\pi} .
	\end{eqnarray}
	Besides, we know that
	\vspace{-.5em}
	\begin{eqnarray}\label{markov_prop2}
	\sum_{i=1}^{N_{sv}+2}\pi_i=1,
	\end{eqnarray}
	where \(\pi_i\) is the \(i\)th index of \(\overrightarrow{\pi}\). By solving the equations of (\ref{markov_prop1}) and (\ref{markov_prop2}), and substituting (\ref{A_N_def}) and (\ref{pi_Un}) in the solution, we can derive the application level availability as
	
	\begin{equation}\label{Pdown}
	\begin{split}
	A=1-\pi_\mathcal{D}&=1-\frac{\bar{\tau}^{-1}_{\mathcal{U}_N}(1-\bar{\tau}^{-1}_{\mathcal{D}_N})^{N_{sv}}} {\bar{\tau}^{-1}_{\mathcal{D}_N}+\bar{\tau}^{-1}_{\mathcal{U}_N}}\\
	&=1-p(1-\bar{\tau}^{-1}_{\mathcal{D}_N})^{N_{sv}}.
	\end{split}
	\end{equation}
	We note that the solution of the proposed FSMC for network level availability results in the same equation as in (\ref{A_N_def}).
	
	As mentioned in section \ref{Reliability_Quantities}, the application level reliability, \(R\), is the mean time for which the communication service is in state \(\mathcal{U}\). Hence, \(R\) can be derived as
	\begin{eqnarray}\label{Rpre}
	R=\frac{A}{L_z},
	\end{eqnarray}
	where \(L_z\) denotes the rate of transitions (i.e., the number of transitions per time unit) to state \(\mathcal{U}\). Accordingly, \(\frac{1}{L_z}\) is a mean time during which two consecutive transitions to state \(\mathcal{U}\) occur. In other words, multiplying \(\frac{1}{L_z}\) with \(A\) results in the mean time period that application is available. The rate of transitions to state \(\mathcal{D}\), \(L_z\), can be calculated as
	\begin{eqnarray}\label{Lz}
	L_z=\pi_{\mathcal{D}}(\bar{\tau}^{-1}_{\mathcal{D}_N})=\frac{\bar{\tau}^{-1}_{\mathcal{U}_N}\bar{\tau}^{-1}_{\mathcal{D}_N}(1-\bar{\tau}^{-1}_{\mathcal{D}_N})^{N_{sv}}} {\bar{\tau}^{-1}_{\mathcal{D}_N}+\bar{\tau}^{-1}_{\mathcal{U}_N}},
	\end{eqnarray}
	where \(\pi_{\mathcal{D}}\) denotes the steady-state probability of state \(\mathcal{D}\). Finally, by substituting (\ref{Lz}) and (\ref{A_N}) in (\ref{Rpre}), we can derive the application level reliability as
	\begin{equation}\label{R}
	\begin{split}
	R=\frac{1}{p\bar{\tau}^{-1}_{\mathcal{D}_N}(1-\bar{\tau}^{-1}_{\mathcal{D}_N})^{N_{sv}}}-\bar{\tau}_{\mathcal{D}_N}.
	\end{split}
	\end{equation}
	In equations (\ref{Pdown}) and (\ref{R}), $\bar{\tau}_{\mathcal{D}_N}$ keeps track of burst error length. Assuming that the failures in the network are independent,  the probability of success and failure does not depend on the current network state. Hence, the probability of failure becomes $p=\bar{\tau}^{-1}_{\mathcal{U}_N} = 1-\bar{\tau}^{-1}_{\mathcal{D}_N}$, and our model is reduced to the simplified independent failure model derived in \cite{Huawei2018}. We evaluate this assumption in the next section and show that it could lead to over-optimistic and inaccurate approximations. 
	
	%
	\vspace{-.1em}
	\section{Simulation Results}
	\vspace{-.1em}
	\begin{figure}
		\vspace{-1.5em}
		\centering
		\includegraphics[width=0.97\columnwidth]{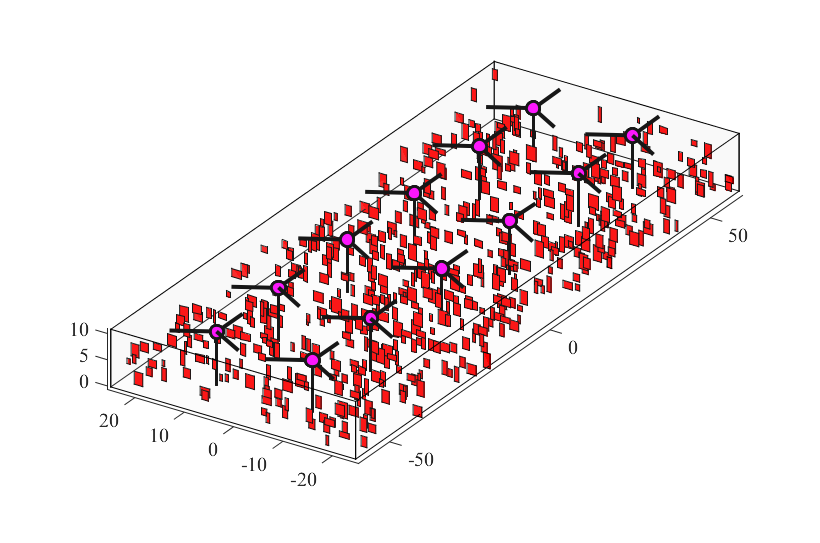}
		\vspace{-1em}
		\caption{Factory model with base stations in black and blockers in red}
		\label{fig:Deployment}
		\vspace{-1.5em}
	\end{figure}
	\begin{table}[htb!]
		\centering
		\caption{Simulation Setup}
		\label{tab:tabSimSetup}
		\begin{tabular}{ |p{3cm}||p{4cm}|  }
			\hline
			\multicolumn{2}{|c|}{Simulation Parameters} \\
			\hline
			Parameter& Value\\
			\hline
			Deployment   & 12 gNBs, 3 cells per gNB\\
			Blocker width   & [0.5, 2] m\\
			Blocker height   & [1, 3] m\\
			Blocker's density   & 0.1 blocker/m$^2$\\
			BS antenna height&   10 m \\
			UE height &1.5 m \\
			Carrier frequency & 4 GHz \\
			Bandwidth&   40 MHz\\
			Subcarrier spacing& 30 KHz  \\
			TTI length & 7 os \\
			Number of BS antennas & 8 \\
			DL transmit power &   27 dBm  \\
			DL noise figure & 9 dB \\
			User lifetime & infinite \\
			Number of Transmissions & 1 \\
			Simulation time &  600 seconds\\
			UE speed & 3 km/h, rail movement \\
			Scheduling   & Round robin, random frequency start point, Dynamic Grant\\
			\hline
		\end{tabular}
		\vspace{-1.5em}
	\end{table}
	
	\begin{figure*}[htb!]
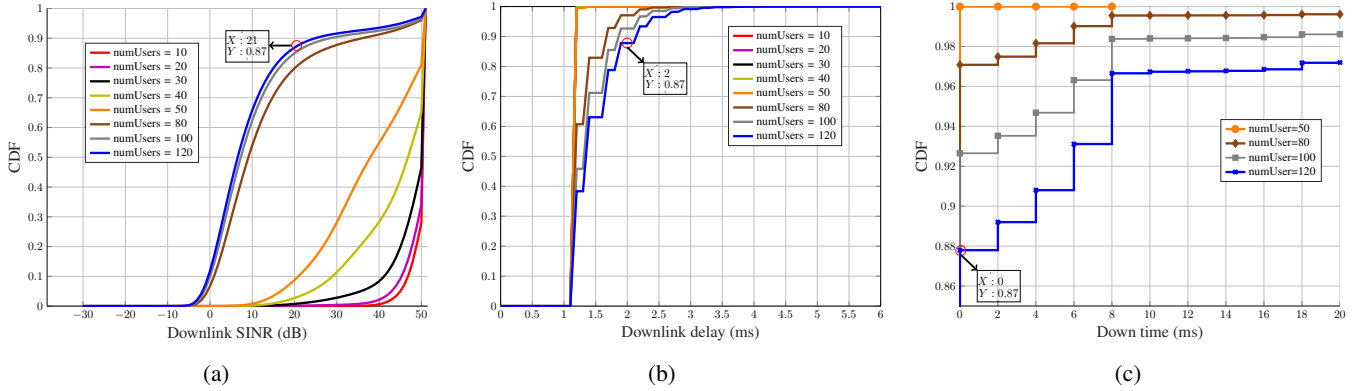

		\begin{subfigure}[b]{0.31\textwidth}
			\centering 
			\scalebox{0.44}{\input{dlSinrCdf.tex}}
			\caption{}
			\label{fig:dlSinrCdf}
		\end{subfigure}
		\hfill
		\begin{subfigure}[b]{0.33\textwidth}
			\centering 
			\scalebox{0.44}{\input{dlDelayCdf.tex}}
			\caption{}
			\label{fig:dlDelayCdf}
		\end{subfigure}
		\hfill
		\begin{subfigure}[b]{0.34\textwidth}
			\centering 
			\scalebox{0.44}{
%
%
\definecolor{mycolor1}{rgb}{0.75000,0.00000,0.75000}%
\definecolor{mycolor2}{rgb}{0.75000,0.75000,0.00000}%
\definecolor{mycolor3}{rgb}{0.55000,0.27000,0.07000}%
\definecolor{mycolor4}{rgb}{0.50196,0.50196,0.50196}%
\begin{tikzpicture}

\begin{axis}[%
width=4.521in,
height=3.565in,
at={(0.758in,0.482in)},
scale only axis,
xmin=0,
xmax=20,
xlabel style={font=\Large \color{white!15!black}},
xlabel={Down time (ms)},
ymin=0.85,
ymax=1,
ylabel style={font=\Large \color{white!15!black}},
ylabel={CDF},
axis background/.style={fill=white},
axis x line*=bottom,
axis y line*=left,
xmajorgrids,
ymajorgrids,
legend style={at={(0.686,0.415)}, anchor=south west, legend cell align=left, align=left, draw=white!15!black}
]
\addplot[const plot, color=orange, line width=2.0pt, mark size=2.5pt, mark=*, mark options={solid, fill=orange, orange}] table[row sep=crcr] {%
0	0\\
0	0.99995892942743\\
2	0.999999776830892\\
4	0.999999990945406\\
6	0.999999998934754\\
8	1\\
};
\addlegendentry{numUser=50}

\addplot[const plot, color=mycolor3, line width=2.0pt, mark size=2.5pt, mark=diamond*, mark options={solid, fill=mycolor3, mycolor3}] table[row sep=crcr] {%
0	0\\
0	0.970807786047175\\
2	0.974898441958341\\
4	0.981596764263295\\
6	0.990176942647898\\
8	0.995496913978641\\
10	0.995547562858366\\
12	0.995566405082339\\
14	0.995599652419223\\
16	0.995736378281339\\
18	0.996127620655453\\
20	0.99614665564961\\
};
\addlegendentry{numUser=80}

\addplot[const plot, color=mycolor4, line width=1.5pt, mark size=2pt, mark=square*, mark options={solid, fill=mycolor4, mycolor4}] table[row sep=crcr] {%
0	0\\
0	0.926509492502061\\
2	0.935246139670915\\
4	0.946832538757847\\
6	0.963150537378733\\
8	0.98377593684611\\
10	0.984036797365417\\
12	0.984118105621077\\
14	0.984236994325027\\
16	0.984636252457041\\
18	0.986072373343479\\
20	0.986136743865322\\
};
\addlegendentry{numUser=100}

\addplot[const plot, color=blue, line width=2.0pt, mark size=2.5pt, mark=x, mark options={solid, fill=blue, blue}] table[row sep=crcr] {%
0	0\\
0	0.877970715084038\\
2	0.892012183220231\\
4	0.908016999422457\\
6	0.93113788342451\\
8	0.966559019366414\\
10	0.967343958390152\\
12	0.967555479574757\\
14	0.967796197105332\\
16	0.968554797418703\\
18	0.971824705613399\\
20	0.97197383074241\\
};
\addlegendentry{numUser=120}

\end{axis}

\begin{axis}[%
width=5.833in,
height=4.375in,
at={(0in,0in)},
scale only axis,
xmin=0,
xmax=1,
ymin=0,
ymax=1,
axis line style={draw=none},
ticks=none,
axis x line*=bottom,
axis y line*=left,
legend style={legend cell align=left, align=left, draw=white!15!black}
]
\draw[red, thick] (1.95cm,2.9cm) circle (1.5mm);
\draw[black, ultra thick, ->] (1.95cm,2.775cm) to (2.45cm,2.3cm); 
\draw[black, thick] ((2.45cm,1.35cm) rectangle (3.75cm,2.3cm);
\filldraw 
(2.85cm,2.225cm) circle (0.1pt) node[align=right,   below] {$X:0$}
(3.1cm,1.37cm) circle (0.1pt) node[align=right,  above] {$Y:0.87$};
\end{axis}
\end{tikzpicture}%
}
			\caption{}
			\label{fig:mdtDist}
		\end{subfigure}
		\vspace{-0.5em}
		\caption{The CDF of (a) DL SINR over PDSCH, (b) network packet delay, and (c) down times for different number of users in the factory}
		\label{fig:dlCdf}
		\vspace{-1.5em}
	\end{figure*}
	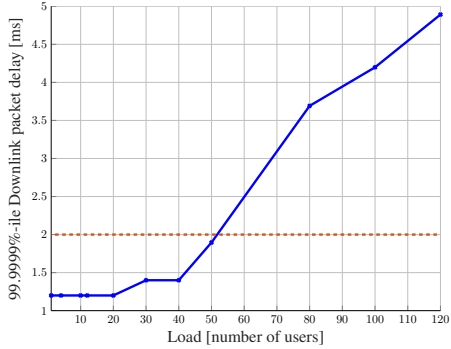
\begin{figure}
		\vspace{-1em}
		\centering 
		\scalebox{0.45}{
%
%
\definecolor{mycolor1}{rgb}{0.85000,0.32500,0.09800}%
\begin{tikzpicture}

\begin{axis}[%
width=11.5cm,
height=9cm,
at={(0.758in,0.519in)},
scale only axis,
xmin=1,
xmax=120,
xlabel style={font=\Large \color{white!15!black}},
xlabel={Load [number of users]},
ymin=1,
ymax=5,
ylabel style={font=\Large \color{white!15!black}},
ylabel={99.9999\%-ile Downlink packet delay [ms]},
axis background/.style={fill=white},
axis x line*=bottom,
axis y line*=left,
xmajorgrids,
ymajorgrids,
legend style={legend cell align=left, align=left, draw=white!15!black}
]
\addplot [color=blue, line width=2.0pt, mark=x, mark options={solid, blue}]
  table[row sep=crcr]{%
1	1.1999999\\
4	1.1999999\\
10	1.19999990033289\\
12	1.19999990027741\\
20	1.19999990033289\\
30	1.39977936654808\\
40	1.39994480426743\\
50	1.89576117431324\\
80	3.69124613395623\\
100	4.19804808652671\\
120	4.89125574604177\\
};

\addplot [color=mycolor1, dashed, line width=2.0pt, forget plot]
  table[row sep=crcr]{%
0	2\\
120	2\\
};
\end{axis}

\begin{axis}[%
width=5.833in,
height=4.375in,
at={(0in,0in)},
scale only axis,
xmin=0,
xmax=1,
ymin=0,
ymax=1,
axis line style={draw=none},
ticks=none,
axis x line*=bottom,
axis y line*=left,
legend style={legend cell align=left, align=left, draw=white!15!black}
]
\end{axis}
\end{tikzpicture}%
}
		\setlength{\abovecaptionskip}{10pt}
		\vspace{-0.5em}
		\caption{The 99.9999 percentile delay for DL PDCP PDUs }
		\label{fig:dlDelayPerc}
		\vspace{-1em}
	\end{figure}
	\vspace{-0.5em}
	In this section, we present the results from the simulations performed to validate the proposed analytical model, and highlight the impact of the application layer's survival time on reliability quantities. Such an impact is significant because of two reasons: (1) it could potentially ease the requirements on network, and hence influences the network design; and (2) it enables us to perform accurate trade-offs between application layer reliability and network resource utilizations. The latter may result in the change of network configuration.
	\subsection{Simulation Methodology}
	We perform both link level and network level simulations. The link level simulation is used to setup the simulation environment, and calculate path gains and 3D channel data. The network level simulations are performed to simulate MAC and higher layers for multiple base stations and users.
	
	In this work, we made an effort to formulate a generic scenario that captures common characteristics of industrial environments, such as large indoor halls, which are cluttered with variable size metallic objects, and surrounded by concrete or metal floors, ceilings, and walls. The factory scenario in \figurename\,\ref{fig:Deployment} is used in this paper. The factory is modeled as a rectangular shape, medium sized plant with dimensions of $120\times50\times11$ m$^3$. We use the 3GPP Indoor Hotspot (InH) model, derived in \cite{3GPP38901} for an office scenario. However, we made adjustments to make it applicable for our large open factory, with concrete walls, ceiling, and floor where there exist many metallic objects. We also introduced a blockage model to capture blocking from humans and other stationary or moving objects. The blockers are rectangular metallic objects of different sizes. We assumed that $ 80\% $ of the blockers are located on the height between $0.5$m to $5$m, and the rest are located on the height between $5$m to $9$m. The height and width of blockers are uniformly distributed in the range of 1m to 3m, and 0.5m to 2m, respectively. To calculate the attenuation caused by each of the blockers, we adopted the blockage model B from \cite{3GPP38901}. In this model, each blocker acts as a spatial filter, and the resulting attenuation on each subpath is derived by a knife-edge diffraction method. In our simulations, the instantaneous SINR leads to an error probability, which is impacted by the radio channel (e.g., path loss and blockage) and the dynamic interference of other transmissions.
	
	In the network level simulations, the radio units are located in two rows with a height of $10$m, where each unit is $20$m away from its neighboring units. The number of radio units is set to 12, where the most left and right ones are $15$ and $10$m away from factory long and short walls, respectively. Each radio unit has $8$ antennas, vertically positioned on top of each other, and pointing horizontally with a down tilting that is optimized to improve the system capacity. On the other hand, the users representing communication nodes on robots are randomly chosen with discrete uniform distribution from the below set (the numbers are in meters),
	\begin{equation*}
	\begin{split}
	x_i &\in \{-58,-56,-54,\dots,56,58\},\\
	y_i &\in \{-24,-22,\dots,22,24\},\\
	\end{split}
	\end{equation*}
	The user's traffic is considered to be from the motion control use case, selected from \cite{3GPP22104}, i.e., cyclic traffic with $2$ms period, delay bound of $2$ms, and the packet size of 20 Bytes. Table \ref{tab:tabSimSetup} summarizes the simulation parameters.
	\subsection{Performance Evaluation}	
	\begin{figure*}
		\begin{minipage}{0.48\textwidth}
			\centering
			\includegraphics[width=0.9\columnwidth]{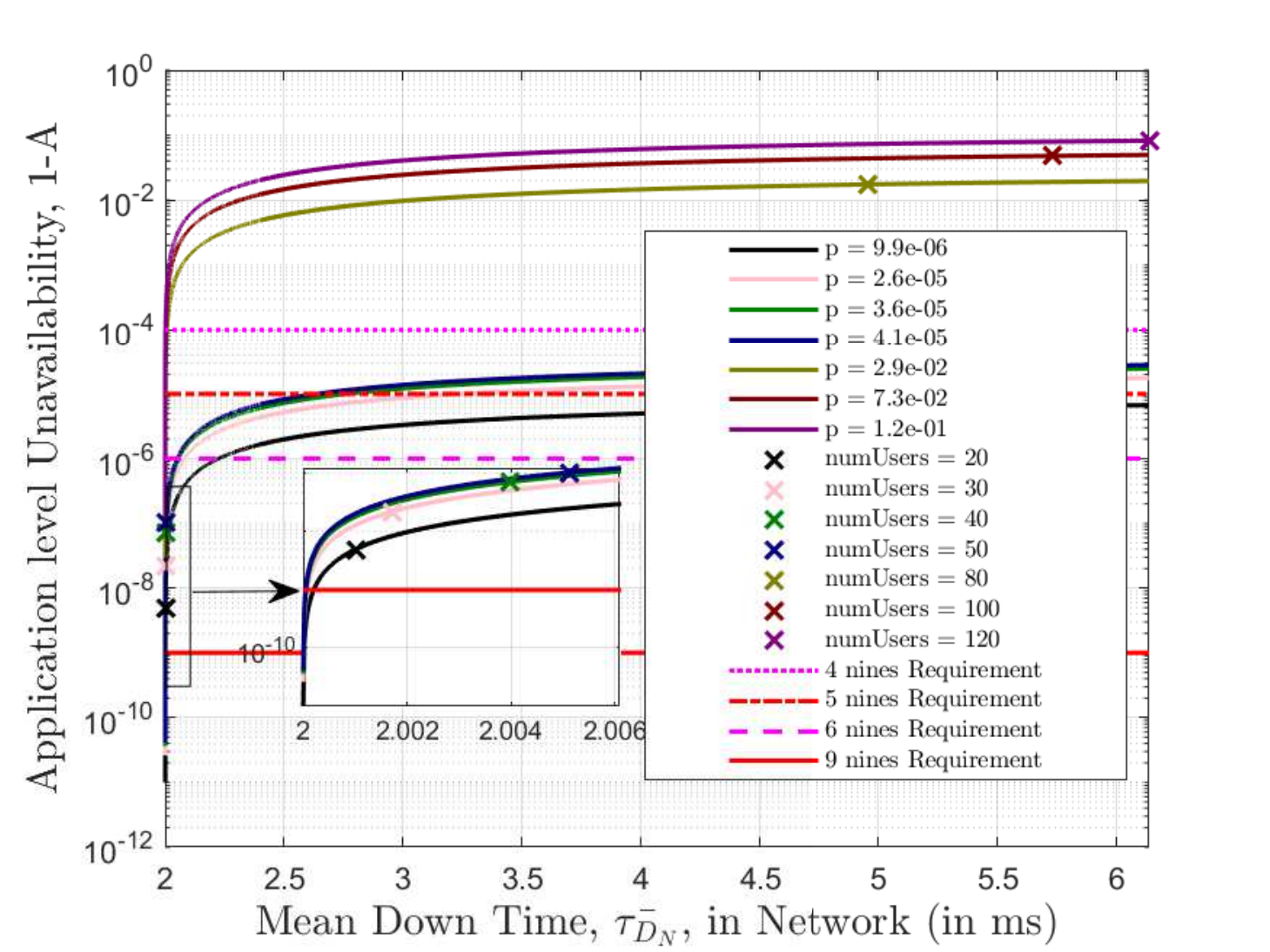}
			\caption{Application layer availability as a function of $\bar{\tau}_{\mathcal{D}_N}$, derived from simulation and (\ref{Pdown}), for different load scenarios}
			\label{fig:aMdt}
		\end{minipage}
		\hfill
		\begin{minipage}{0.48\textwidth}
			\centering
			\includegraphics[width=0.9\columnwidth]{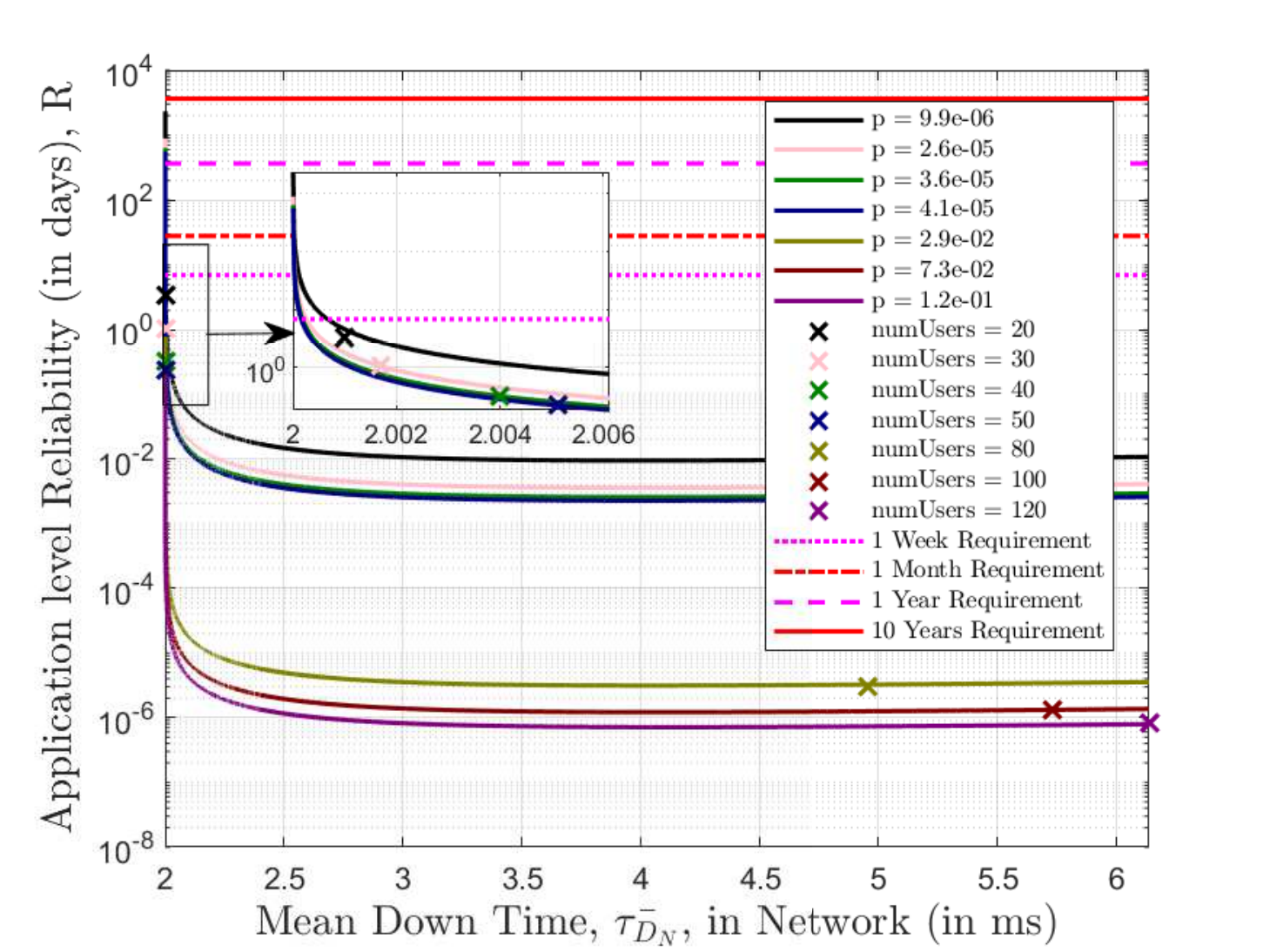}
			\caption{Application layer reliability derived from simulation and (\ref{R}), as a function of $\bar{\tau}_{\mathcal{D}_N}$ for different load scenarios}
			\label{fig:rMdt}
		\end{minipage}
		\vspace{-1.5em}
	\end{figure*}
	
	For our evaluations, we measured different performance metrics on the user side, while solely considering downlink (DL) traffic. We ran simulations while iterating over the number of users in the factory to evaluate our proposed mapping function under various load conditions, i.e., the higher number of users, the more load in the network. Each simulation point with a specific number of users was run for at least 100 times. Here, we first present more general performance measures, namely signal to interference plus noise ratio (SINR) and DL delay, to get a better understanding of the network characteristics. Then, we evaluate our reliability quantities and compare them to the developed theoretical results.
	
	To derive reliability and availability metrics, we store the following for each round of simulation:
	\begin{itemize}
		\item  $TX_{i,j}$: the sequence number for the $i$th transmitted PDCP PDU to user $j$
		\item $RX_{i,j}$: the sequence number for the $i$th received PDCP PDU in user $j$
		\item $D_{i,j}$: the observed delay of $RX_{i,j}$.
	\end{itemize}
	We consider a PDU sent by user $j$ as lost if it is either not found in the $RX_{i,j}$, or if its corresponding delay, $D_{i,j}$, is more than its delay bound. This procedure results in a binary matrix, $\mathbf{B}$, where $1$ represents a PDU that is successfully and timely received, and $0$ represents unsuccessful or delayed reception. Hence, given the assumption of cyclic traffic, the packet error ratio, $p$, and $\bar{\tau}_{\mathcal{D}_N}$ can be derived at the PDCP layer. On the other hand, for calculating the application layer metrics, the concept of survival time needs to be applied. Hence, depending on the defined $N_{sv}$  on each scenario, we replace up to $N_{sv}$ consecutive $0$s, happening right after a transition from $1$, with $1$s in matrix $\mathbf{B}$. This results in reduction or disappearance of  $\tau_{D_{N,i}}$s. The derived matrix is the sequence observed by the application layer.
	
	\figurename\,\ref{fig:dlCdf}(\subref{fig:dlSinrCdf}) shows the CDF of the SINR on the physical downlink shared channel (PDSCH). The CDF is calculated over all samples measured in all simulation runs for each network load. \figurename\,\ref{fig:dlCdf}(\subref{fig:dlDelayCdf}) illustrates the CDF of DL delays measured on the PDCP layer. This delay is the measured latency of protocol data unit (PDU) from the transmitting PDCP entity to the receiving PDCP entity, and hence, considers solely the successfully received PDUs. According to \figurename\,\ref{fig:dlCdf}(\subref{fig:dlSinrCdf}), there is a significant jump, up to 0.7, in SINR CDF when the number of users increases from $50$ to $80$, implying that the probability of an arbitrary SINR sample being less than a threshold increases significantly.  Correspondingly, a similar behavior happens on the DL delay distribution, meaning that the DL PDCP PDUs experience a noteworthy decrease in terms of latency when the number of users rises from $50$ to $80$.  \figurename\,\ref{fig:dlCdf}(\subref{fig:mdtDist}) illustrates the empirical CDF of down times evaluated on PDCP layer. In this figure, each probability on y axis represents the proportion of packets that are either successfully and timely received, or part of a down time which is equal or less than the corresponding value on x axis. For example, when there are $120$ users in the factory, the probability that a packet is part of a down time of less than $6$ms is $0.91$, and the probability that it is part of up time is $0.87$. The CDF of $0.87$ corresponds to $2$ms DL delay, in \figurename\,\ref{fig:dlCdf}(\subref{fig:dlSinrCdf}), and 21 dB SINR, in \figurename\,\ref{fig:dlCdf}(\subref{fig:dlSinrCdf}), implying that, in this example, delay bound is the major cause of the experienced packet loss.
	
	\figurename\,\ref{fig:dlDelayPerc} shows the 99.9999 percentile delay of the successfully received PDCP PDUs. It can be seen that the 99.9999 percentile delay in $80$-user scenario is around 1.8 ms higher than the delay in $50$-user scenario. Since our simulations perform single shot transmissions (i.e., no retransmissions), this increase could be caused by the queuing delay introduced with higher loads.
	\begin{figure*}[hbt!]
		\begin{subfigure}[b]{0.49\textwidth}
			\centering
			\includegraphics[width=0.9\columnwidth]{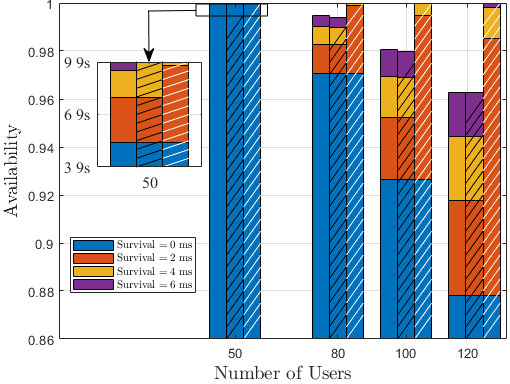}
			\caption{}
			\vspace{-.5em}
			\label{fig:barA}
		\end{subfigure}
		\begin{subfigure}[b]{0.49\textwidth}
			\centering
			\includegraphics[width=0.9\columnwidth]{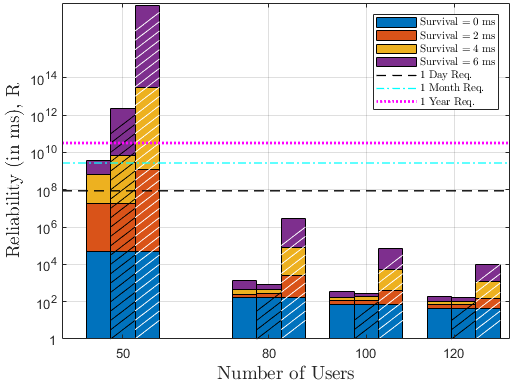}
			\caption{}
			\vspace{-.5em}
			\label{fig:barR}
		\end{subfigure}
		\caption{Impact of surival time on application layer's (a) availability, and (b) reliability, based on simulation results, plain colored on the left side, our theoretical results, black hatched in the middle, and estimations under the assumption of independent failures, white hatched on the right side.}
		\label{fig:bar}
		\vspace{-1.5em}
	\end{figure*} 
	
	\figurename\,\ref{fig:aMdt} shows the application layer unavailability, $1-A$, as a function of $\bar{\tau}_{\mathcal{D}_N}$, when the number of consecutive failures that can be tolerated by application layer (i.e., $N_{sv}$), is $1$. In this figure, the crosses show the simulation results, and the lines represent the theoretical results. For the calculation of the latter, we use $p$ from the simulation and plot the equation (\ref{Pdown}) when iterating over different $\bar{\tau}_{\mathcal{D}_N}$. The figure shows clearly that our method estimates the application layer availability with high precision. In \figurename\,\ref{fig:rMdt}, we illustrate the application layer reliability based on $\bar{\tau}_{\mathcal{D}_N}$ where $N_{sv}=1$. Similar to the previous figure, the crosses show the simulation result while the lines are derived from (\ref{R}), where $p$ is the PDCP packet error ratio from the simulation. Simulation results here show an acceptable match with the theoretical approximations. In both of the figures, there is a jump in $p$ when the number of users increases from $50$ to $80$. This jump could be well motivated by the SINR reduction in \figurename\,\ref{fig:dlCdf}(\subref{fig:dlSinrCdf}) and DL delay increase in \figurename\,\ref{fig:dlCdf}(\subref{fig:dlDelayCdf}), when comparing $50$-user load with $80$-user load. In \figurename\,\ref{fig:aMdt} and \figurename\,\ref{fig:rMdt}, we also plot the typical requirements, specified in \cite{3GPP22104}, for availability and reliability, respectively.  Our results show that, for the selected configuration, the availability requirement of six 9s (i.e., $99.9999\%$) could be fulfilled with a low number of users. However, there is still a big gap between the system reliability in our configuration and reliability requirements for motion control traffic specified in \cite{3GPP22104}.
	
	\figurename\,\ref{fig:bar} illustrates the impact of survival time on availability and reliability for different selected load scenarios on the network. Left bars, plain colored, are derived from our simulations, the middle ones, hatched with black, are derived from  (\ref{Pdown}), for availability, and (\ref{R}), for reliability, and the right ones, hatched with white, are derived with the assumption of independent packet failures (for derivation, refer to  \cite{Huawei2018}). As can be seen in \figurename\,\ref{fig:bar}(\subref{fig:barA}), in $50$-user scenario, simulations show that the application layer availability can fulfill the eight 9s (i.e., $0.99999999$) availability requirement when the survival time is three times the cycle time. The results suggest that survival time can impact availability to a significant extent. For instance, in the $80$-user scenario, the availability of $97\%$ can increase to up $99.4\%$ when $N_{sv}=3$, which could be sufficient for applications with looser requirements (e.g., plant asset management). \figurename\,\ref{fig:bar} shows that the assumption of independent failures can lead to extreme optimism on the approximation of the application performance. For example, in \figurename\,\ref{fig:bar}(\subref{fig:barA}), the approximation based on independent failures for $80$-user scenario with $T_{sv}$ of $2$ms,  results in 0.999 availability (e.g., sufficient for positioning traffic); however, the actual availability derived by simulations shows that the availability, in this case, is below $0.99$. The results in \figurename\,\ref{fig:bar}(\subref{fig:barA}), \figurename\,\ref{fig:bar}(\subref{fig:barR}) shows that our model offers a decent approximation of the application level performance, except for $50$-user load when the survival time is more than $4$ms. After a careful analysis of the results, we concluded that our simulation length contributed to this difference between simulation and theoretical results. For the $50$-user scenario, we ran 600-second simulations for 250 times, i.e., a total of less than 87 days of simulation time, however, it is not possible to validate the reliability of more than ten years with this kind of discrete event based simulations where excessive details result in long run time.
	
	\section{Conclusion}
	Modern cyber-physical control applications can tolerate sparse packet loss. In this paper, we applied fundamental reliability theory metrics into wireless communication scenarios and derived a function which maps application level performance to network level parameters, and vice versa. In addition to packet error ratio, our model incorporates burst errors and provides this mapping given the target application specification on survival time. The benefits of our model are twofold. First,  in the process of designing the network, it provides the end-to-end performance of the proposed deployment, implying that the network designer can approximate the application level performance without the need to perform computationally complex simulations to validate years of reliability. Second, in run time, it enables the network controller to adjust the trade-off between network reliability and utilization based on the current approximation of application level performance. We validated our theoretical results with detailed simulations of the industrial automation scenario. We showed that the assumption of independent failures tends to an optimistic estimation of application level performance, resulting in unmet application demands.

	
	%
	
	\vspace{-1em}
	\appendices
	%

	%
	%

	\ifCLASSOPTIONcaptionsoff
	\newpage
	\fi

	
	
	
	\bibliographystyle{IEEEtran}
	\bibliography{IEEEabrv,ref}
	\vspace{-1em}
	%
\end{document}